\documentclass{article}

\usepackage{graphicx} 
\usepackage{appendix}
\usepackage{amsmath}
\usepackage{amssymb}
\usepackage{float}

\usepackage{blindtext}
\usepackage{geometry}
\usepackage{url}
\newcommand{\bd}{\begin{displaymath}}
\newcommand{\ed}{\end{displaymath}}
\newcommand{\be}{\begin{equation}}
\newcommand{\ee}{\end{equation}}
\newcommand{\bea}{\begin{eqnarray}}
\newcommand{\eea}{\end{eqnarray}}
\newcommand{\bda}{\begin{eqnarray*}}
\newcommand{\eda}{\end{eqnarray*}}
\newcommand{\ba}{\begin{array}}
\newcommand{\ea}{\end{array}}
\newcommand{\ddd}{\mbox{\,{\scriptsize \rm d}}}
\newcommand{\R}{{\bf R}}
\newcommand{\N}{{\bf N}}
\newcommand{\Rn}{{\bf R}^n}
\newcommand{\Rr}{\R^r}
\newcommand{\Rm}{\R^m}
\newcommand{\Ra}{\Rightarrow}
\newcommand{\ra}{\rightarrow}
\newcommand{\Mt}{\Rightarrow}
\newcommand{\mt}{\mapsto}
\newcommand{\st}{\subset}
\newcommand{\lra}{\longrightarrow}
\newcommand{\Lra}{\Longrightarrow}
\newcommand{\sm}{\setminus}
\newcommand{\A}{{\bf A}}
\newcommand{\B}{{\bf B}}
\newcommand{\Bo}{{\stackrel{\circ}{{\bf B}}}}
\newcommand{\LL}{{\cal L}}
\newcommand{\U}{{\cal U}}
\newcommand{\V}{{\cal V}}
\newcommand{\G}{{\cal G}}
\newcommand{\F}{{\cal F}}
\newcommand{\M}{{\cal M}}
\newcommand{\C}{{\cal C}}
\newcommand{\SSS}{{\cal S}}
\newcommand{\di}{\mbox{\rm dist}}
\newcommand{\E}{{\cal E}}
\newcommand{\T}{{\cal T}}
\newcommand{\Q}{{\cal Q}}
\newcommand{\HH}{{\cal H}}
\newcommand{\e}{\varepsilon}
\newcommand{\al}{\alpha}
\newcommand{\ph}{\varphi}
\newcommand{\sth}{ : \,}
\newcommand{\de}{\stackrel{def}{=}}
\newcommand{\dd}{\mbox{\rm\,d}}
\newcommand{\cl}{\mbox{\rm cl}\,}
\newcommand{\gr}{\mbox{\rm graph}\,}
\newcommand{\crn}{\mbox{\rm comp}(\Rn)}
\newcommand{\co}{\mbox{\rm co}\,}
\newcommand{\h}{\mbox{\rm haus}}
\newcommand{\cn}{\mbox{\rm cone}\,}
\newcommand{\ir}{\mbox{\rm int}\,}
\newcommand{\Lm}{\mathop{\mbox{\rm Lim}}\limits}
\usepackage{authblk}
\title{Pattern Formation in a Spiking Neural-Field of Renewal Neurons}
\date{}
\author[1]{Gr\'{e}gory Dumont}
\author[2,3]{Carmen Oana Tarniceriu}
\affil[1]{School of Life Sciences, Ecole Polytechnique F\' ed\' erale de Lausanne, Switzerland}
\affil[2]{Department of Mathematics and Informatics, "Gheorghe Asachi" University of Ia\c si, Ia\c si, Romania,}
\affil[3]{The Institute of Interdisciplinary Research, Department of Exact Sciences and Natural Sciences,“Alexandru Ioan Cuza” University of Ia\c si, Ia\c si, Romania.}
\begin{document}
\maketitle

\begin{abstract} Elucidating the neurophysiological mechanisms underlying neural pattern formation remains an outstanding challenge in Computational Neuroscience. In this paper, we address the issue of understanding the emergence of neural patterns by considering a network of renewal neurons, a well-established class of spiking cells. Taking the thermodynamics limit, the network's dynamics can be accurately represented by a partial differential equation coupled with a nonlocal differential equation. The stationary state of the nonlocal system is determined, and a perturbation analysis is performed to analytically characterize the conditions for the occurrence of Turing instabilities. Considering neural network parameters such as the synaptic coupling and the external drive, we numerically obtain the bifurcation line that separates the asynchronous regime from the emergence of patterns. Our theoretical findings provide a new and insightful perspective on the emergence of Turing patterns in spiking neural networks. In the long term, our formalism will enable the study of neural patterns while maintaining the connections between microscopic cellular properties, network coupling, and the emergence of Turing instabilities.
\end{abstract}

{\bf Keywords:}
Spiking neural networks, Neural fields models, Age-Structured equations, Pattern formation
\\


\section{Introduction}
Pattern formation is a ubiquitous phenomenon in nature, evident across diverse systems \cite{ball}. The initial exploration of pattern formation in biological systems traces back to Alan Turing's groundbreaking reaction-diffusion theory. Turing patterns, or Turing instabilities, represent a self-organizing spatial pattern that emerges in reaction-diffusion systems — a concept introduced in his seminal 1952 paper \cite{turing}.
Since Turing's pioneering work, his ideas have permeated various fields within developmental biology \cite{maini1,maini2,othmer} and have been proposed as explanations for a multitude of biological phenomena \cite{munteanu,murray1,newman}.

This concept of pattern formation holds particular significance in Computational Neuroscience, where neural field models have become standard tools for studying such phenomena. Initially introduced by Beurle \cite{beurle} and expanded upon by Wilson and Cowan \cite{W_C1,W_C2}, the model has evolved into its current standard form introduced by Amari \cite{A00,A01}. The neural field equations, expressed as integro-differential equations, capture the non-linear dynamics of large populations of neurons. The exploration of pattern formation using neural field models was  pioneered by Amari \cite{A01}, who extended the application of continuum models of neural activity to pattern formation. This extension was made under natural assumptions about connectivity and firing rate functions. In his work, Amari considered local excitation and distal inhibition, effectively modelling a mixed population of interacting inhibitory and excitatory neurons with typical cortical connections. The nonlinearity and range of interactions between neurons are crucial elements in understanding pattern formation in neural fields.

Neural field models have provided significant additional value to our understanding of the emergence of patterns in the brain. They offer a clear explanation of visual hallucinations \cite{bressloff_h, ermentrout}, binocular rivalry \cite{binocular}, orientation tuning in the visual cortex \cite{orientation} or the formation of ocular dominance columns in the visual cortex \cite{miller}. For a comprehensive overview, interested readers may consult the following review papers on the topic: \cite{Bres01, Bres_stoc}. We also recommend that the reader consult the textbooks \cite{Ermentrout_2, Bres02} for an intuitive introduction to this topic.  

While neural field models have significantly contributed to our understanding, it is crucial to acknowledge their inherent limitations in capturing the full spectrum of neural activity. They only provide a rate-based approximation, i.e., they only model the average firing rate of neurons in a population, rather than the individual spikes of each neuron. On another hand, spiking neural networks provide a biologically realistic representation of the discrete firing events of individual neurons, which is essential for understanding neural synchronization and complex activity patterns. The introduction of noise into such networks further enhances their fidelity to real neural systems, as it mirrors the inherent variability and unpredictability of firing events.  
In response to this challenge, stochastic neural field models emerged as a promising avenue \cite{Bres_stoc, touboul1,touboul2}, offering a nuanced perspective on the role of randomness in phenomena like wave regulation \cite{kilpatrick1} or wandering bumps \cite{kilpatrick2}, or in inducing bifurcations and shaping the onset of pattern formation \cite{carillo_pat1,carillo_field1}.

As research in neural modelling progressed, attention shifted towards adapting these models to spiking networks. This evolution led to the derivation of neural field models specifically tailored for theta neurons \cite{luke1} and quadratic integrate-and-fire  neurons \cite{montrbio1}. Subsequent extensions, incorporating spatial dependencies \cite{laing1,laing2}, enriched the modelling framework. These adapted spiking network models have proven instrumental in analyzing a diverse range of phenomena, including Turing instabilities, synchrony-induced modes of oscillations, and the existence and stability of spatio-temporal patterns \cite{byrne2}. Finite size effects have been studied \cite{shotnoise} and they have also been applied to the study of synchrony-induced modes of oscillations \cite{acebes1}, as well as of the emergence of bumps and oscillations within spiking networks \cite{schmidt_avitabile}. Moreover, their utility extends to electroencephalography modelling \cite{byrne}.

As evidenced by this overview, analytical treatments are emerging   and hold the promise of understanding the emergence of pattern formation in spiking circuits. 
In this paper, we consider a spiking network where a given cell is characterized by 
the amount of time passed by since its last action potential, i.e. 
the age of the cell. 

Among the ways to describe spiking neural networks, stands out the refractory density approach \cite{GH}. In this framework, the neurons are described by a variable designating the time that elapsed since their last spike \cite{perthame03}. The model consists in a partial differential equation with a non-local boundary condition, very well known in the literature as the age-structured population model.
Age-structured systems, a modelling approach that tracks the distribution of individuals across different age groups, have proven to be invaluable tools in understanding dynamic processes in various scientific disciplines. Originating in the early 20-th century, with pivotal contributions from Sharpe, Lotka \cite{Lotka}, and McKendrick \cite{mckendrick}, these models have evolved in the form mostly used today introduced by VanFoerster \cite{foerster}. The age-structure formalism have developed to become indispensable in capturing essential features of real-world data. Its adaptability is showcased in epidemiology \cite{epid1,epid2}, in cellular proliferation studies \cite{cell1}, shedding light on population dynamics at the microscopic level, and in broader ecological contexts.


However, it is in the realm of neuroscience that the age-structured formalism has recently found particularly intriguing applications. By extending beyond traditional population dynamics, age-structured models offer a nuanced perspective on neuronal activity, providing insights into phenomena such as the locking mechanisms of excitatory networks \cite{g2000}, emergence of synchronized activity \cite{g2000} or transient dynamics \cite{deger} and the dynamic behaviour of finite-size effects \cite{schw_gerst, SchwDeger,greg3, meyer}. The age-structured approach has become an invaluable tool in unravelling the complex dynamics of neural systems, furthering our understanding of the brain's intricate functioning. While there can be other ways to express the dynamics of spiking neural networks, age structure formalism has its advantages. It has been shown that is amenable to describe renewal processes such as the noisy integrate and fire \cite{usdt,usit,usmmnp} or  spike-response models \cite{g2000}, low-dimensional reductions \cite{SchwPietras}, and shares connections with conductance based models \cite{ch1,ch2}.  The interested reader may refer to the review \cite{tiloschw} on this topic. Compared to other models describing spiking neural networks, the age-structured models are simple enough in form to be more mathematically tractable.   From this point of view, analytical properties of solutions have been proven, including the construction of periodic analytical solutions \cite{perthame04}, reduction to delay equations \cite{salort21}, and generalization to cases that include the time elapsed since the penultimate discharge \cite{salort22}.
 Properties of spatially distributed networks of Spike Response Models neurons have been analyzed in \cite{gerstner}(chapter 9), where a range of phenomena such as travelling or periodic waves are represented.
 Recently, a  model of age structured neural network incorporating spatial dependence and learning processes has been also considered in \cite{TS2020}, where authors analyze the convergence towards  stationary states in the case of weak connections along with the situation when the learning process is faster than the activity of the network. Moreover, the model can be derived from  stochastic processes taking place at microscopic level \cite{Chev1}.

This manuscript aims to elucidate the synaptic mechanisms underlying the emergence of global Turing patterns in a network of spiking neurons. 
 We achieve this by introducing a novel continuity equation, combining age-structured and neural field formalisms. To this end, we consider a network of renewal neurons uniformly distributed on a periodic domain.  In the present paper, we leverage the mean-field formalism to reduce the network dynamics to a single nonlocal partial differential equation. The resulting continuity equation captures processes occurring at both the cellular and network levels. It is a hybrid of the age-structured equation and the neural fields equation. 
To the best of our knowledge, this is the first time that such a continuity equation has been introduced.

We then compute the steady state of the equation, which gives us access to the average activity in the asynchronous regime. By linearizing around the asynchronous solution, we use perturbation analysis to compute the bifurcation diagram, which predicts the emergence of Turing instabilities in the nonlocal continuity equation. We illustrate our theoretical findings with numerical simulations, and systematically compare simulations of the continuity equation with those of the fully stochastic network. For parameters below or above the bifurcation line, we show that the simulated spiking activity of the full network confirms the emergence of a transition from an asynchronous to a Turing type of neural activity. 

This paper is structured as follows. First, we introduce the network and neuron model that will be used throughout. Next, we introduce the nonlocal partial differential equation that describes the neural network and perform a linear stability analysis to investigate the emergence of Turing patterns. We finish the paper with a discussion regarding possible perspectives. To facilitate the reading, we let the mathematical details for the appendix section.

\section{Spiking Neural Network and Mean-Field Description}

We consider a network of  $N$ neurons uniformly  distributed in a periodic spatial domain  $[-\pi,\pi]^n$, where $n$ denotes the dimension of the space in which we consider the domain. Assuming that the cells in the network are described  by renewal processes, then, for a given cell, only the time since its last action potential accounts for describing the cell's dynamics. Denoting the total input received  by a neuron $h_N(t,x)$,  the probability of spiking of a neuron placed in position $x$ during a time interval $[t,t+dt]$ is be given by $S(h_N(t,x),r)\dd t$, where $r$ is the time elapsed since the last spike of the cell.

In this setting, it is considered that after a spike is triggered, the neuron's age $r$ is reset to zero.
The total activity at instant $t$  at position $x$ is then given by the sum of all the occurring spikes:
\begin{equation}
\label{FR_N}
A_{N}(t,x)= \frac{1}{N}  \sum_{k=1}^{N} \sum_{f} \delta (t-t_k^f) \delta(x-x_k ),  
\end{equation}
where $\delta$ is the Dirac mass, $t_k^f$ the firing time of the cell numbered $k$ and positioned in $x_k$.
The total input current received by the neurons in position $x$ is given by 
\bd
h_N(t,x)= I_{ext}(t,x)   + I_N(t,x),
\ed
where $I_{ext}(t,x)$ is an external current and the synaptic current $I_N(t,x)$, which defines the current feedback of the network, is given by

\begin{figure}[t!]
\begin{center}
    \includegraphics[width=\textwidth]{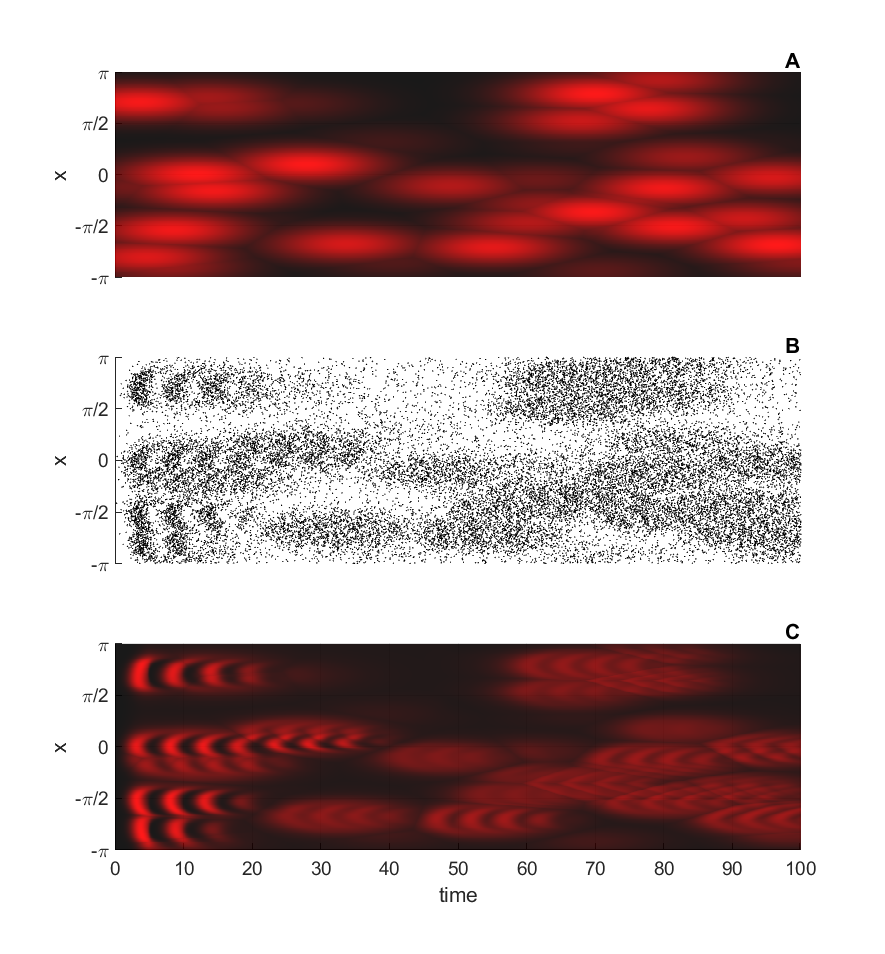}
       \caption{Comparison of the firing rate activity between the full network and its corresponding mean-field equation for the ring network. {\bf A.} A space and time dependent external current $I_{ext}(t,x)$. {\bf B.} The raster plot and the firing rate extracted from Monte-Carlo  simulations of the full network containing $2500$ neurons, described by the model (\ref{FR_N}). {\bf C.} The firing rate given by the simulation of the probability density function (\ref{FR}). Parameters: $T=5$, $\tau=5$, $J_s=5$, $\alpha=1/2$. }
       \label{Figure4}
      \end{center}
\end{figure}

\be\label{syn_cN}
\tau \frac{\partial }{\partial t} I_N(t,x)=-I_N(t,x) +\int\displaylimits_{ [-\pi,\pi]^n } J(x-y)A_N(t,y)\dd y.
\ee
Here $\tau$ is the synaptic time constant and $J$ is the connectivity function, which, in our case, will only depend on the
distance between neurons placed in positions $x$ and $y$. Furthermore, the connectivity function is assumed to be periodic on the spatial domain $[-\pi,\pi]^n$.

Taking now the thermodynamic limit  ( i.e. when $N$ goes to infinity), the full network description reduces to a single partial differential equation, endowed with a nonlocal boundary condition that describes the reset mechanism. 
Denoting by $q(t,r,x)$ the probability density for a neuron in position $x$ to have at time $t$ the age $r$, the density profile evolves according to the continuity equation:
\begin{equation}\label{AS}
\frac{\partial}{\partial t}q(t,r,x) +\frac{\partial}{\partial r}q(t,r,x)=-S(h(t,x),r)q(t,r,x).
\end{equation}
Because once a cell emits an action potential its age is reset to zero, the
natural boundary condition is
\bd
q(t,0,x)=A(t,x),
\ed
where $A(t,x)$ is the activity at moment $t$ in the spatial position $x$ and is extracted from
\begin{equation}
\label{FR}
A(t,x)= \int_{0}^{+\infty}S(h(t,x),r)q(t,r,x) \dd r .
\end{equation}

\begin{figure}[t!]
\begin{center}
    \includegraphics[width=\textwidth]{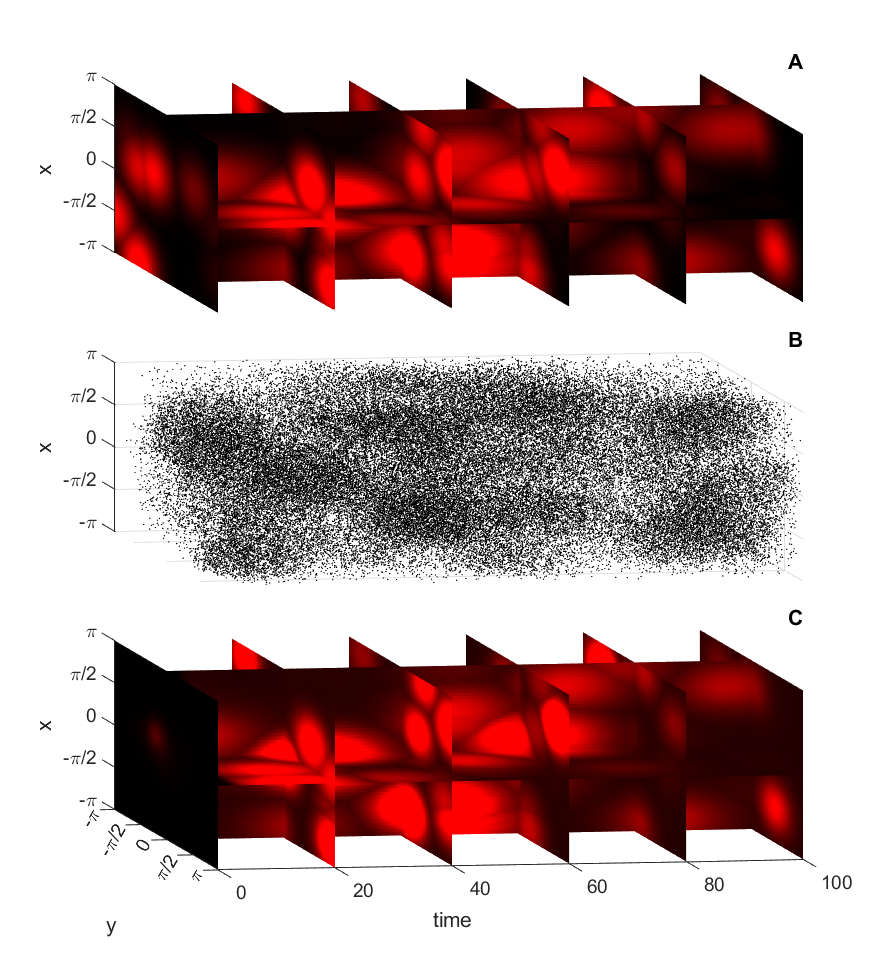}
       \caption{Comparison of the firing rate activity between the full network and its corresponding mean-field equation for the torus network. {\bf A.} A space and time dependent external current $I_{ext}(t,x)$. {\bf B.} The raster plot and the firing rate extracted from Monte-Carlo simulations of the full network of 12500 neurons described by the model (\ref{FR_N}). {\bf C.} The firing rate given by the simulation of the probability density function (\ref{FR}). Parameters: $T=5$, $\tau=5$, $J_s=5$, $\alpha=1/4.1$.  See the supplementary material for different views of the figure.}
       \label{Figure5}
      \end{center}
\end{figure}

We stress that in the thermodynamic limit the total input current is given now by 
\bd
h(t,x)= I_{ext}(t,x) + I(t,x),
\ed
with
\be\label{syn_c}
\tau \frac{\partial }{\partial t} I(t,x)=-I(t,x) +\int\displaylimits_{ [-\pi,\pi]^n } J(x-y)A(t,y)\dd y.
\ee
We will furthermore consider spatial periodicity at the boundary of the domains.
The initial density is assumed known:
\bd
q(0,r,x)=q_0(r,x).
\ed
Finally, the mean-field equation must obey a conservation law, 
therefore 
\bd
 \int_0^\infty q(t,r,x)\dd r =(2 \pi)^{-n}, \forall t>0, \forall x \in [-\pi,\pi]^n,
\ed
as soon as
\bd
\int_0^\infty q_0(r,x)\dd r=(2 \pi)^{-n}.
\ed
 This assumption is based on the natural hypothesis that neurons are distributed homogeneously across the spatial domain. 
Therefore,  $q(t,x,r)$ defines a probability density function in the sense:
\bd
\int\displaylimits_{ [-\pi,\pi]^n } \int_0^\infty q(t,r,x)\dd r\dd x=1, \quad \forall t>0.
\ed
As stated above, we assume spatial periodicity, and therefore we consider $J$ defined on the domain $[-\pi,\pi]^n$ to be a periodic function. To construct such a synaptic kernel function, we define

\bd
J(x)=J_s\sum_{l_1=-\infty}^{+\infty}...\sum_{l_n=-\infty}^{+\infty} w\left(\left(\sum_{k=1}^n(x_k+2\pi l_k)^2\right)^\frac{1}{2}\right), \quad x=(x_1,..,x_n)\in [-\pi,\pi]^n
\ed
where $w$ is a scalar function, see appendix for details. Importantly, here the coefficient $J_s$ reflects the synaptic efficiency and its effect will be studied in the course of this paper.

Although the theory and simulations' results hold for a wide variety of functions, in the whole paper the spiking probability will be given by 
$$S(h(t,x),r)= e^{h(t,x)}H(r-T),$$ 
with $H$ the Heaviside function, and the connectivity will be taken as $$w(t)=e^{- |t|}- \al e^{- \frac{|t|}{2}}.$$


Finding an analytical solution of the mean-field equation is exceedingly difficult, if not impossible. However, information about the dynamics of the neural activity  can be obtained via numerical simulations. In Fig \ref{Figure4}, a numerical simulation of the fully connected network on a ring ($n=1$) is displayed. In the first panel, Fig \ref{Figure4}{\bf A}, the time course of a given external current is shown. The corresponding spiking activity of the full network and the firing rate extracted from it (\ref{FR_N}) can be observed in Fig \ref{Figure4}{\bf B}. In Fig \ref{Figure4}{\bf C}, the time course of the firing rate given by the density approach (\ref{FR}) is presented. 

Similarly, in Fig \ref{Figure5},  numerical simulations of the fully connected network on a torus ($n=2$)   are presented. In the first panel, Fig \ref{Figure5}{\bf A}, the time course of a given - spatially distributed -   external current is shown. The corresponding spiking activity of the full network of neurons and the firing rate extracted from  (\ref{FR_N}) can be observed in Fig \ref{Figure5}{\bf B}.
In Fig \ref{Figure5}{\bf C}, the time course of the firing rate given by the density approach (\ref{FR}) is presented.  Fig \ref{Figure5}{\bf A} and 
\ref{Figure5}{\bf C} illustrate a continuous in time representation of the 2D distributions of the external current and activity intersected with planes at different instants of time.



As one can see, the mean-field equation effectively captures  the essential features   of the full neural activity for both the ring and torus network.
In what follows, we shall discuss the case of one-dimensional space distribution $(n=1)$, respectively the two-dimensional case $(n=2)$, and the interest to find under which conditions the network loses stability and Turing patterns may arise. To do so, we will start by discussing the asynchronous state, and continue with the stability analysis of the stationary solution by means of perturbation method. 
\begin{figure}[t!]
\begin{center}
    \includegraphics[width=\textwidth]{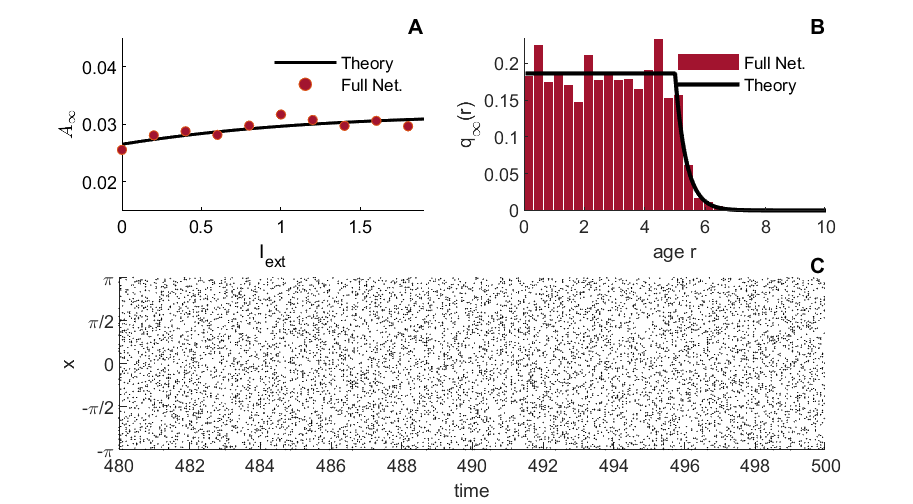}
   \caption{The asynchronous state solution of the ring network. In panel {\bf A}, a comparison of the stationary activity obtained analytically and Monte Carlo simulations of the full network is presented.  In panel {\bf B}, the analytical stationary state is plotted as the black line, while the histogram  presents the corresponding density extracted from simulations of the full network of $20000$ neurons. In panel {\bf C}, a raster plot of the activity of the full network in the last $20$ ms of the simulations is extracted. One can notice the convergence toward the uniformly distributed steady state. Parameters: $T=5$, $\tau=5$, $J_s=5$, $\alpha=1/2$. }
   \label{Figure9}
      \end{center}
\end{figure}

\section{The Asynchronous State}

\begin{figure}[t!]
\begin{center}
    \includegraphics[width=\textwidth]{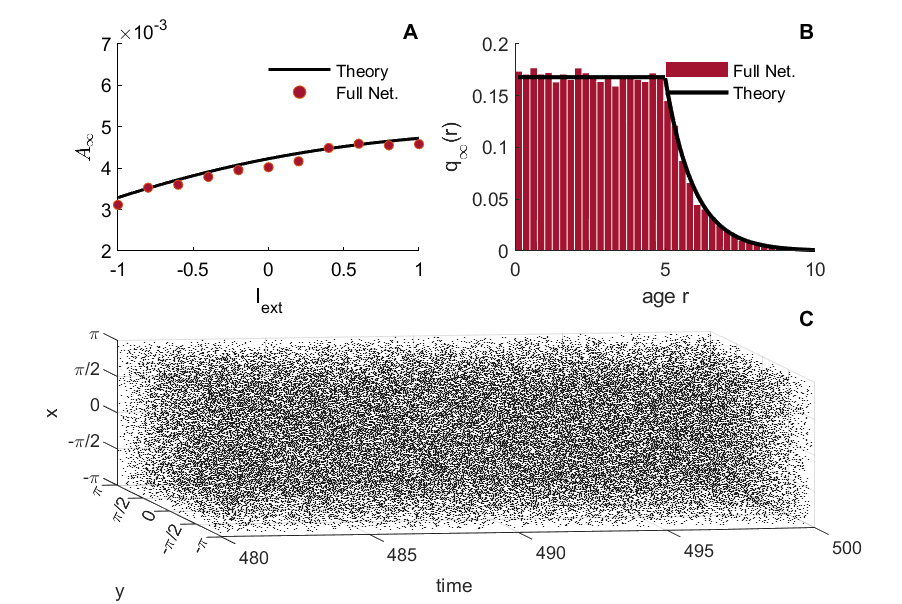}
   \caption{The asynchronous state solution of the torus network. In the panel {\bf A}, a comparison of the stationary activity obtained analytically and Monte Carlo simulations of the full network containing $25000$ neurons is presented. In panel {\bf B}, the analytical stationary state is plotted as the black line, while the histogram  presents the corresponding density extracted from simulations of the full network. In panel {\bf C}, a raster plot of the activity of the full network in the last $20$ ms of the simulations is extracted. One can notice the convergence toward the uniformly distributed steady state. Parameters: $T=5$, $J_s=50$, $\tau=5$, $\alpha=1/4.1$. }
   \label{Figure8}
      \end{center}
\end{figure}
\begin{figure}[t!]
\begin{center}
    \includegraphics[width=\textwidth]{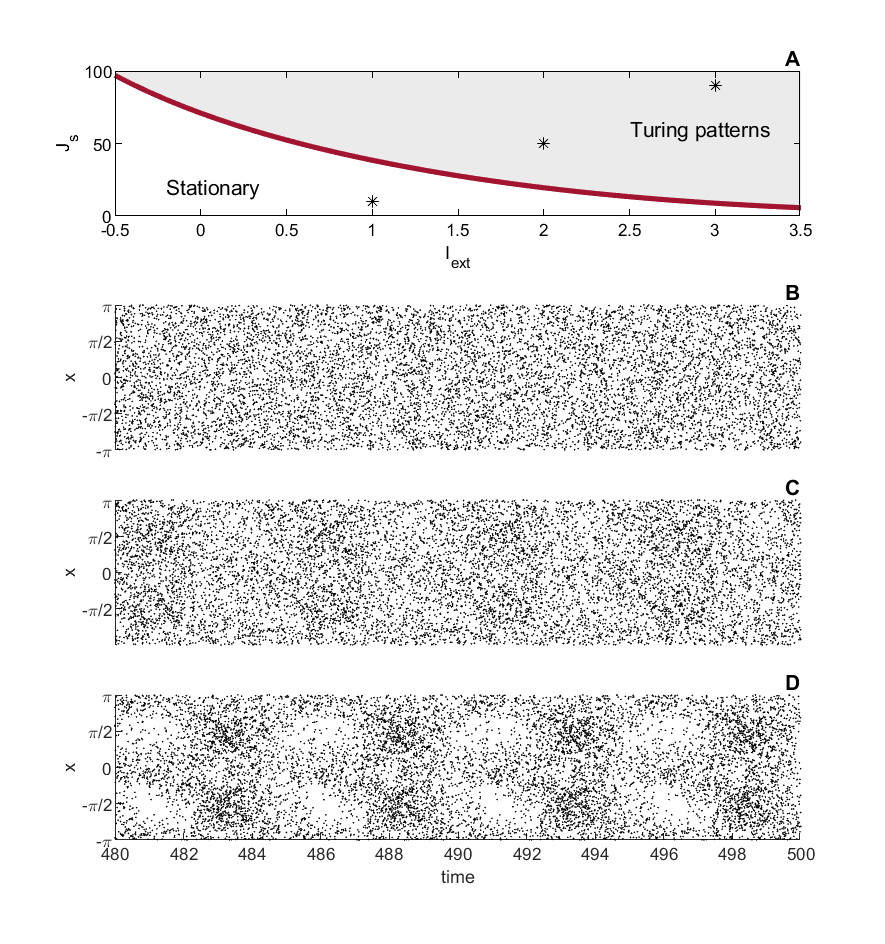}
   \caption{Pattern formation on the ring network. In panel {\bf A}, the bifurcation line that separates the asynchronous state from a Turing pattern regime given by (\ref{CE}), is shown. The panels {\bf B}, {\bf C}, {\bf D} present the corresponding activities extracted from simulations of the full network  of $2500$ neurons, for the three sets of parameters depicted in panel {\bf A}, namely $(I_{ext}, J_s)=(1,10)$, $(I_{ext}, J_s)=(2,50)$ and $(I_{ext}, J_s)=(3,90)$. In panel {\bf B}, the convergence toward the steady state is observed. In panel {\bf C},  the onset of Turing patterns can be spotted, while in panel {\bf D} the patterns in the activity are more obvious. Parameters: $\tau =5$, $T=5$, $\alpha=1/2$. }
   \label{Figure1}
      \end{center}
\end{figure}

The asynchronous state can be computed as the time independent solution of the mean-field equation.
We are looking for a time independent solution for the above system, and also a space independent solution, which will imply a constant mean firing rate over time and space, solution to the following equation (see Appendix for details of the computations performed for obtaining the results in this section):
 Let us denote $q_{\infty}(r)$ the steady state, and $A_{\infty}$ the mean firing rate. 
By direct integration, the stationary solution is obtained in the form:
\be\label{ss}
q_{\infty}(r)= A_{\infty} e^{- \int_0^r  S(h_{\infty},s) \ddd s  }.
\ee
Finally, the asynchronous mean firing rate can be computed using the conservation property of the neural network

 \be\label{SFR}
\left(( 2\pi)^n A_{\infty}\right)^{-1}  = \int_0^{\infty} e^{- \int_0^r  S(h_{\infty},s) \ddd s  }   \dd r.
\ee

Note that the mean firing rate is only implicitly given, since $h_{\infty}$ does depend on $A_{\infty}$. Indeed, we have
\be
h_{\infty} = I_{ext} + A_{\infty} \int\displaylimits_{ [-\pi,\pi]^n } J(y)\dd y.
\ee

\begin{figure}[t!]
\begin{center}
    \includegraphics[width=\textwidth]{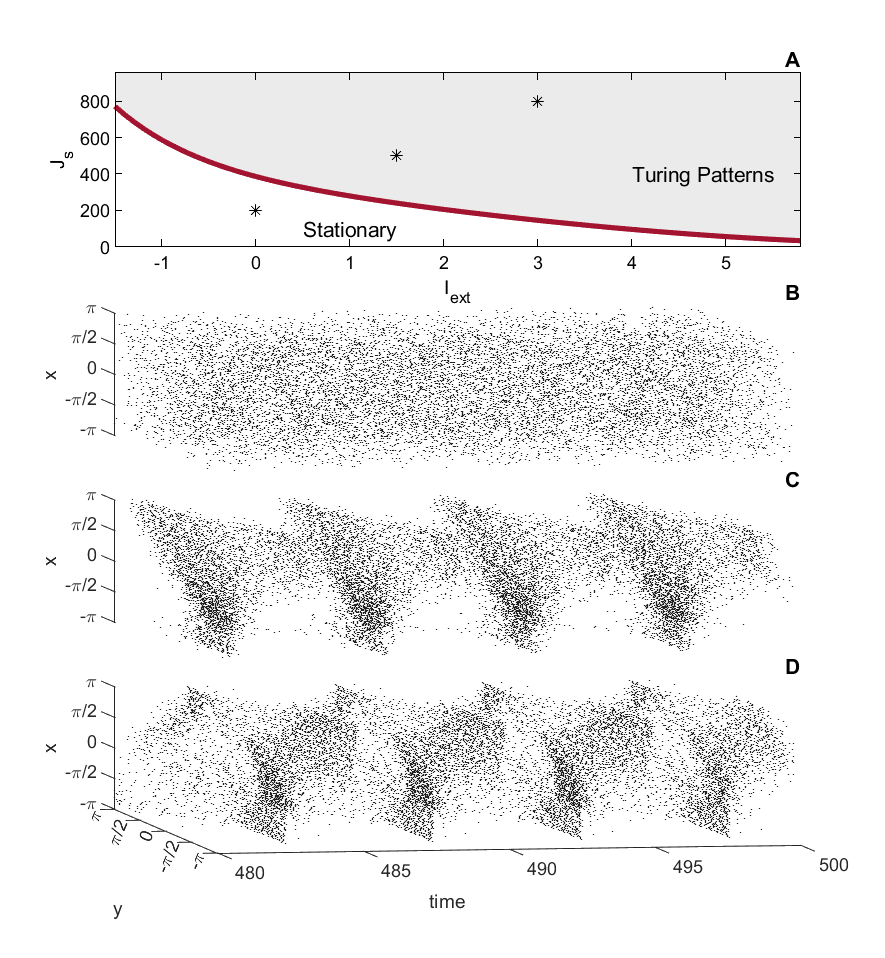}
       \caption{Pattern formation on the torus network. In the panel {\bf A}, the bifurcation line that separates the asynchronous state from a Turing pattern regime given by (\ref{CE}), is shown. The panels {\bf B}, {\bf C}, {\bf D} present the corresponding activities extracted from simulations of the full network for the three sets of parameters depicted in panel {\bf A}, namely $(I_{ext}, J_s)=(0,200)$, $(I_{ext}, J_s)=(1.5,500)$ and $(I_{ext}, J_s)=(3,800)$. In panel {\bf B}, the convergence toward the steady state is observed. In panel {\bf C},  the onset of Turing patterns is already clearly observed. Panel {\bf D} presents  the activity of the network for a pair of parameters situated farther from the bifurcation line, where, again, the Turing patterns are obvious. Parameters: $T=5$, $\tau=5$, $\alpha=1/4.1$. See the supplementary material for different views of the figure.}
       \label{Figure7}
      \end{center}
\end{figure}

The equation above must be solved self-consistently, see Appendix for details. From Fig \ref{Figure9} and Fig \ref{Figure8}, the agreement between analytical and numerical results is excellent.
In figures Fig \ref{Figure9} and Fig \ref{Figure8}, panels {\bf A},  comparisons between the stationary firing rate and by the simulation of the whole network is shown, along with the extracted firing rates Fig in panels {\bf C} for the ring and torus network, respectively. Panels {\bf B} compare the analytical solution and a histogram of age density extracted from the simulation of the whole network.

\section{Turing patterns}

To study the stability of the asynchronous state, one needs to look at the eigenvalues of the differential operator once a linearization around the steady state has been performed. To this end, we consider small perturbations of the steady state by writing the solution $q$  in the form 

\bd
q(t,r,x) = q_{\infty}(r)+ \varepsilon e^{\lambda t + ikx}q_1(r) + o(\varepsilon ^2),
\ed
and similarly for the activity 
\bd
A(t,x) = A_{\infty}+ \varepsilon e^{\lambda t + ikx}A_1 + o(\varepsilon ^2).
\ed
 This allowed us to obtain the characteristic equation, whose solutions give the eigenvalues $\lambda$ of the deterministic system associated to the $k$ Fourier coefficient. The characteristic equation is in the form

\be\label{CE}
\mathcal{C}(\lambda,k)=0,
\ee
where

\bda
\mathcal{C} (\lambda,k) &=&      1-\hat J (k) \kappa_{\lambda}  \int_{0}^{+\infty}\frac{\partial S}{\partial h}(h_{\infty},r)q_{\infty}(r)\dd r -  \int_{0}^{+\infty}S_{\infty}(r) e^{- \int_0^{r}  S(h_{\infty},s)  +\lambda  \dd s } \dd r \\
&&-\hat J (k) \kappa_{\lambda}      \int_{0}^{+\infty}S(h_{\infty},r) \int_0^{r} \frac{\partial S}{\partial h} (h_{\infty},a)q_{\infty}(a) e^{- \int_a^{r}  S(h_{\infty},s)  +\lambda  \ddd s } \dd a \dd r.
\eda

Above,
\bd
 \kappa_{\lambda} = \int_0^{\infty} \kappa (s)\exp (- \lambda s) \dd s,
\ed
where
\bd
\kappa (t)=\frac{1}{\tau}e^{-\frac{t}{\tau}},
\ed
see the Appendix section for more details.\\
The eigenvalues of the linearized operator are thus given by the roots of this equation. The time-independent solution  will be stable if all eigenvalues have negative real parts. The bifurcation line  — which separates the oscillatory regime from the asynchronous one  - can be drawn numerically. 
The important parameter in the bifurcation is given by $\hat J (k)$, with $k\in\mathbb{Z}^n$, $n=1,2$, which is the $k$ Fourier coefficient of the periodic coupling function:

\be
\hat J (k) = \int\displaylimits_{ [-\pi,\pi]^n } J(x) e^{-ikx}\dd x.
\ee
We describe in the appendix how the Fourier coefficient can be computed using the Poisson summation formula and the Bessel function.
 Following standard procedures, one can determine the bifurcation line for the maximum value of $\hat{J}$, which is well-defined since $J$ is symmetric, making the Fourier modes real numbers. For this maximum value, at least one Fourier mode loses stability.

In Fig. \ref{Figure1}, the line separating the stationary and Turing patterns regimes with respect to parameters $I_{ext}$ and $J_s$ can be depicted in panel {\bf A}. In panels {\bf B} and {\bf C}, the corresponding activity to the two sets of parameters shown in panel {\bf A} is plotted. Panel {\bf B} shows the steady state, while in panel {\bf C}, one can already observe emergence of patterns.
Farther from the bifurcation line, the patterns of the network's activity become more visible as it can be seen in Fig.\ref{Figure1}{\bf D} .

Similarly, in Fig. \ref{Figure7}, the line separating the stationary and Turing patterns domains can be depicted in panel {\bf A} for the torus network. In panels {\bf B} and {\bf C}, the corresponding activity to the two sets of parameters shown in panel {\bf A} is plotted. Panel {\bf B} shows the steady state, while in panel {\bf C}, one can already observe emergence of patterns. Once again, further away from the bifurcation line, the patterns of the network's activity become much stronger.

\section{Conclusion}


In unravelling the mysteries of neural function, the significance of pattern formation in neuronal models cannot be overstated. 
From influencing hallucinations \cite{bressloff_h} to guiding spatial orientation \cite{W_C1}, 
patterns play a pivotal role in understanding the various facets of brain  function. 
The traditional way through which neuroscience has approached this complex phenomenon is through neural field models \cite{Bres02,Ermentrout_2}. 
However, these models, while insightful, reveal limitations when tasked with capturing the nuances inherent in spiking networks.

Imported from statistical physics, the mean field approximation has been successfully adapted to describe the dynamics of spiking neural networks. Under the assumption of an infinitely large population of neurons that share the same characteristics, a continuity equation giving the evolution of a probability density function is obtained \cite{DH}. Probably the most well-known equation relied upon in the mean field approach is the Fokker-Planck equation \cite{Gardiner}, but depending on the type of neurons in the population, other formalisms are obtained. Age-structured systems were particularly derived for the case of spike response model neurons \cite{GH}. The method, in particular, lends itself to descriptions of the network dynamics in terms of a population density function \cite{sirovich02, us,DH}. The resulting continuity equation provides the means to mathematically describe a number of phenomena that are not otherwise available. Among these, neural synchronization has been at the core of interest. The phenomenon has been scrutinized by considering key elements thought to induce it: synaptic delay \cite{ikeda}, strength of connectivity \cite{Henry}, interaction of inhibitory and/or excitatory populations \cite{sirovich01}, bistability between an asynchronous state and an oscillatory state \cite{OBH}, the effect of noise \cite {B-H}, the neurons' refractoriness \cite{deger}, or combinations of the elements mentioned above \cite{usjns}. Furthermore, this framework can be employed to study brain oscillations characterization through phase-resetting curves \cite{greg1} and adapted to describe finite-size networks \cite{greg3}. A unified perspective that addresses the issue of neuronal synchronization and pattern formation for spatially distributed neurons would be a step forward in understanding neural dynamics.

In recent years, advances in adapting neural fields models to incorporate the characteristics of spiking neural networks have been made in the case of theta and quadratic integrate-and-fire neurons \cite{montrbio1, luke1}, allowing the analysis of a range of phenomena \cite{shotnoise,schmidt_avitabile,acebes1}. Furthermore, the extension to include spatial dependence has allowed the investigation of occurring Turing instabilities \cite{byrne2}.
Nevertheless, the investigation of pattern formation in spiking networks described by more general models remains an open ended problem.


Recognizing the inherent strengths of both neural field models and mean field models for spiking neurons, we propose on an innovative approach. By  integrating both frameworks, we introduced a model that not only addresses the limitations posed by traditional neural field models but also allows for the simultaneous analysis of pattern formation and bifurcation points. 

While previous efforts have extensively explored the role of synaptic 
connections in Turing-type network activity, our paper presents a theoretical framework for studying emergent macroscopic network patterns in a population model of renewal neurons. Our methodology provides a pathway to explore the intricate relationships between microscopic cellular parameters, network coupling, and the emergence of Turing patterns in the brain. Connectivity strength emerges as a pivotal factor in pattern generation, and we demonstrate how other parameters, such as external current, influence pattern onset in a noisy setting.

Moreover, our model opens avenues for several extensions. 
The synaptic delay is considered a key element in inducing synchronized activity; therefore, adapting our model to include a synaptic delay parameter/function,
which can be embedded  either in the age-structured equation  \cite{greg3}, or in the field equation \cite{SKG}, would be a valuable addition. Another potential extension involves analyzing the effect of adaptation which has been the subject of investigation of the neural field equation \cite{EFK}, and can be incorporated  in the age-structured equation \cite{SchwDeger}. 
Considering finite-size fluctuations is crucial for capturing the effects of noise on the system \cite{greg3}, and it can be interesting to see how would influence the dynamics of our model. On the other hand, the shape of the spatial connectivity function, which influences the patterns' shape, is a feature that can be the subject of more detailed investigations. Additionally, one can always consider separated inhibitory/excitatory populations of neurons and examine how this will affect synchronized activity in spatial patterns. Finally, special solutions such as spiral waves or bump solutions can be analyzed, and a natural extension of our study would be to investigate the existence of travelling waves.

Summarizing, our theoretical framework not only contributes to the ongoing discourse on pattern emergence in neural networks but also offers a versatile platform for exploring the nuanced dynamics of brain function and further extending our understanding of complex biological phenomena.

\appendix
\section{The Mean-Field Approach}


Denoting by $q(t,r,x)$ the probability density for the neuron in position $x$ to have at time $t$ the age $r$, the density profile evolves according to the continuity equation:

\begin{equation}
\frac{\partial}{\partial t}q(t,r,x) +\frac{\partial}{\partial r}q(t,r,x)=-S(h(t,x),r)q(t,r,x) \quad t>0, \quad x\in [-\pi,\pi]^n
\end{equation}
with the boundary condition is
\bd
q(t,0,x)=A(t,x),
\ed
where $A(t,x)$ is given by
\begin{equation}
A(t,x)= \int_{0}^{+\infty}S(h(t,x),r)q(t,r,x) \dd r ,
\end{equation}
and
\bd
h(t,x)= I_{ext}(t,x) + I(t,x),
\ed
with
\be
\tau \frac{\partial }{\partial t} I(t,x)=-I(t,x) +\int\displaylimits_{ [-\pi,\pi]^n } J(x-y)A(t,y)\dd y.
\ee
The initial density is assumed known:
\bd
q(0,r,x)=q_0(r,x).
\ed
To complete the description of the network, the conservation of probability should be imposed,
therefore 
\bd
 \int_0^\infty q(t,r,x)\dd r =(2 \pi)^{-n}, \forall t>0, \forall x \in [-\pi,\pi]^n,
\ed
as soon as
\bd
\int_0^\infty q_0(r,x)\dd r=(2 \pi)^{-n},
\ed
where $n$ is the dimension of the space in which the network is considered to be displaced. Therefore the function $q(t,x,r)$ defines a probability function in the sense:
\bd
\int\displaylimits_{ [-\pi,\pi]^n } \int_0^\infty q(t,r,x)\dd r\dd x=1, \quad \forall t>0.
\ed

\subsection{The asynchronous state}
 Let us denote $q_{\infty}(X,r)$ the steady state, and $A_{\infty}(X) $ the mean firing rate. We will look for a space independent solution, and therefore assume 
 $$q_{\infty}(X,r)=q_{\infty}(r),\quad A_{\infty}(X)=A_{\infty}.$$
Then, the following equation takes place

\bd
\frac{d}{d r}q_{\infty}(r)=-S(h_{\infty},r)q_{\infty}(r),
\ed
where we have denoted 
\be
h_{\infty} = I_{ext} + A_{\infty} \int\displaylimits_{ [-\pi,\pi]^n } J(y)\dd y.
\ee

By direct integration, the stationary solution is obtained in the form:
\bd
q_{\infty}(r)= A_{\infty} e^{- \int_0^r  S(h_{\infty},s) \,ds  },
\ed
where we have used the  boundary condition
\bd
q_{\infty}(0)= A_{\infty}.
\ed
Finally, the asynchronous mean firing rate can be computed using the conservation property of the neural network

\bd
 \int_0^{\infty} q_{\infty}(r) \dd r =(2\pi)^{-n},
\ed
to get
\bd
\frac{1}{ A_{\infty}\left( 2\pi\right)^{n}}  = \int_0^{\infty} e^{- \int_0^r  S(h_{\infty},s) \ddd s  }   \dd r,
\ed
\begin{figure}[t!]
\begin{center}
    \includegraphics[width=\textwidth]{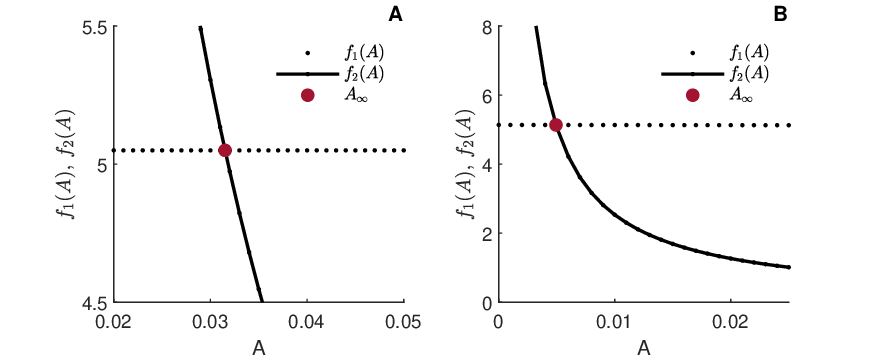}
       \caption{Graphical solutions of the stationary states for the ring network (panel {\bf A}) and torus network (panel {\bf B}). The solutions are obtained as the intersection of the curves  $f_1(A)$ and $f_2(A)$. The analytical solution given by (\ref{ss}) is marked by a red dot in both panels. Parameters: $T=5$, $J_s=5$, $\tau=5$, $\alpha=1/2$ for $n=1$ and  $T=5$, $J_s=50$, $\tau=5$, $\alpha=1/4.1$ for $n=2$.  }
       \label{Figure10}
      \end{center}
\end{figure}
The way of solving the nonlinear equation expressing $A_{\infty}$ is shown in figure Fig \ref{Figure10}{\bf A}. The solution of the stationary firing rate $A_{\infty}$ is obtained as an intersection of the curves:

\be
f_1(A)=\int_0^{+\infty} e^{-\int_0^r S\left(I_{ext} + A \bar J\right)\ddd s}\dd r , \quad f_2(A)=\left( A\left( 2\pi\right)^{n}\right)^{-1} .
\ee
with
$$\bar J := \int\displaylimits_{ [-\pi,\pi]^n } J(y)\dd y$$

\subsection{The characteristic equation}
Let us consider first order perturbations of the steady state solution that we considered space-independent. Then, we assume that
\bd
q(t,r,x) = q_{\infty}(r)+ \varepsilon q_1(t,r,x) + o(\varepsilon ^2)
\ed
and similarly for the activity 
\bd
A(t,x) = A_{\infty}+ \varepsilon A_1(t,x) + o(\varepsilon ^2).
\ed

Plugging these expressions and only keeping the first order term, we get that the perturbation obeys to the following partial differential equation

\be\label{q1}
\frac{\partial}{\partial t}q_1(t,r,x) +\frac{\partial}{\partial r}q_1(t,r,x)=-S_{\infty}(r)q_1(t,r,x)  -  \frac{\partial}{\partial h}S(h_{\infty},r)q_{\infty}(r) I_1 (t,x) ,
\ee
and the perturbation of the activity is given by
\be\label{A1}
 A_1(t,x)=\int_{0}^{+\infty}S(h_{\infty},r)q_1(t,r,x) \dd r + I_1 (t,x)\int_{0}^{+\infty}\frac{\partial}{\partial h}S(h_{\infty},r)q_{\infty}(r)\dd r. 
\ee
Above, we used the fact that, due to the form of the activity $A$, the synaptic current is given by
\bd
I(t,x)=I_\infty+\varepsilon I_1(t,x)+ o(\varepsilon ^2)
\ed
and
\bd
h_\infty=I_{ext}+I_\infty=I_{ext}+A_\infty\int_{[-\pi, \pi]^n} J(y)\dd y.
\ed
Since we are interested by pattern formation, we look for a solution of the form

\bd
 q_1(t,r,x) =e^{i k x}\bar q_1(t,r) ,
\ed
and similarly, an activity given by

\bd
 A_1(t,x) =e^{i k x}\bar A_1(t) ,
\ed
Plugging these expressions in (\ref{q1}), respectively, (\ref{A1}), it follows that

\bd
\frac{\partial}{\partial t} \bar q_1(t,r) +\frac{\partial}{\partial r}\bar q_1(t,r)=-S_{\infty}(r)\bar q_1(t,r)  - \frac{\partial}{\partial h} S(h_{\infty},r)q_{\infty}(r) \hat J (k)\left(\kappa *  \bar A_1\right)(t), 
\ed
where we have used the fact that $I_1(t,x)=\bar I_1(t) e^{i k x}$ with 
\bd
\tau \frac{d}{d t}  \bar I_1(t)=-\bar I_1(t)+\bar A_1(t)\hat J(k).
\ed
Above, $*$ denotes the time convolution,
\bd
\kappa (t)=\frac{1}{\tau}e^{-\frac{t}{\tau}},
\ed
and $\hat J (k) $ is the Fourier transform of the synaptic kernel.
The activity obeys
\bd
 \bar A_1(t)=\int_{0}^{+\infty}S(h_{\infty},r)q_1(t,r)\dd r +  \hat J (k)  \kappa *  A_1(t) \int_{0}^{+\infty}\frac{\partial S}{\partial h}(h_{\infty},r)q_{\infty}(r)\dd r.     
\ed

To analyze the stability of such a solution, one looks for it in the form

\bd
 \bar q_1(t,r) =e^{\lambda t}\tilde q_1(r) ,
\ed
and a similar assumption is made for the neural activity

\bd
\bar A_1(t) = e^{\lambda t} \tilde A_1 .
\ed
By straightforward computations, we get that the perturbation obeys to
\be\label{tildeq1}
\frac{\dd}{\dd r}\tilde q_1(r)=-\left(S(h_{\infty},r)  +\lambda \right) \tilde q_1 (r)   +\hat J (k) \tilde A_1 \frac{\partial S}{\partial h} (h_{\infty},r)q_{\infty}(r)  \kappa_{\lambda},
\ee
where we  have introduced the new notation
\bd
 \kappa_{\lambda} = \int_0^{\infty} \kappa (s)\exp (- \lambda s) \dd s.
\ed
Similarly,
\bd
\tilde A_1\left(1-\hat J (k) \kappa_{\lambda} \int_{0}^{+\infty}\frac{\partial S}{\partial h}(h_{\infty},r)q_{\infty}(r)\dd r \right)  =\int_{0}^{+\infty}S_{\infty}(r)\tilde q_1(r) \dd r.  
\ed
Integrating the linear differential equation (\ref{tildeq1}), we get

\bd
\tilde q_1(r) = \tilde  A_1 e^{- \int_0^{r}  \left(S(h_{\infty},s)  +\lambda\right)  \ddd s }  + \tilde A_1 \hat J (k) \kappa_{\lambda}   \int_0^{r}  \frac{\partial S}{\partial h}(h_{\infty}, a)q_{\infty}(a) e^{- \int_a^{r}  \left(S(h_{\infty},s)  +\lambda\right)  \ddd s } \dd a,
\ed
which implies 

\bda
&&\int_{0}^{+\infty}S(h_{\infty},r)\tilde q_1(r) \dd r =\tilde A_1  \int_{0}^{+\infty}S(h_{\infty},r) e^{- \int_0^{r}  \left(S(h_{\infty},s)  +\lambda\right)  \ddd s } \dd r\\
&&+\hat J (k) \kappa_{\lambda} \tilde A_1   \int_{0}^{+\infty}S(h_{\infty},r) \int_0^{r} \frac{\partial S}{\partial h} (h_{\infty},a)q_{\infty}(a) e^{- \int_a^{r} \left(S(h_{\infty},s)  +\lambda\right)  \ddd s } \dd a  \dd r.
\eda
After some computations, we finally arrive to the following equation

\bda
&&  1-\hat J (k) \kappa_{\lambda}  \int_{0}^{+\infty}\frac{\partial S}{\partial h}(h_{\infty},r)q_{\infty}(r)\dd r -  \int_{0}^{+\infty}S_{\infty}(r) e^{- \int_0^{r}  S(h_{\infty},s)  +\lambda  \ddd s } \dd r\\
&&-\hat J (k) \kappa_{\lambda}      \int_{0}^{+\infty}S(h_{\infty},r) \int_0^{r} \frac{\partial S}{\partial h} (h_{\infty},a)q_{\infty}(a) e^{- \int_a^{r}  S(h_{\infty},s)  +\lambda  \ddd s } \dd a \dd r   =0.
\eda

\begin{figure}[t!]
\begin{center}
    \includegraphics[width=\textwidth]{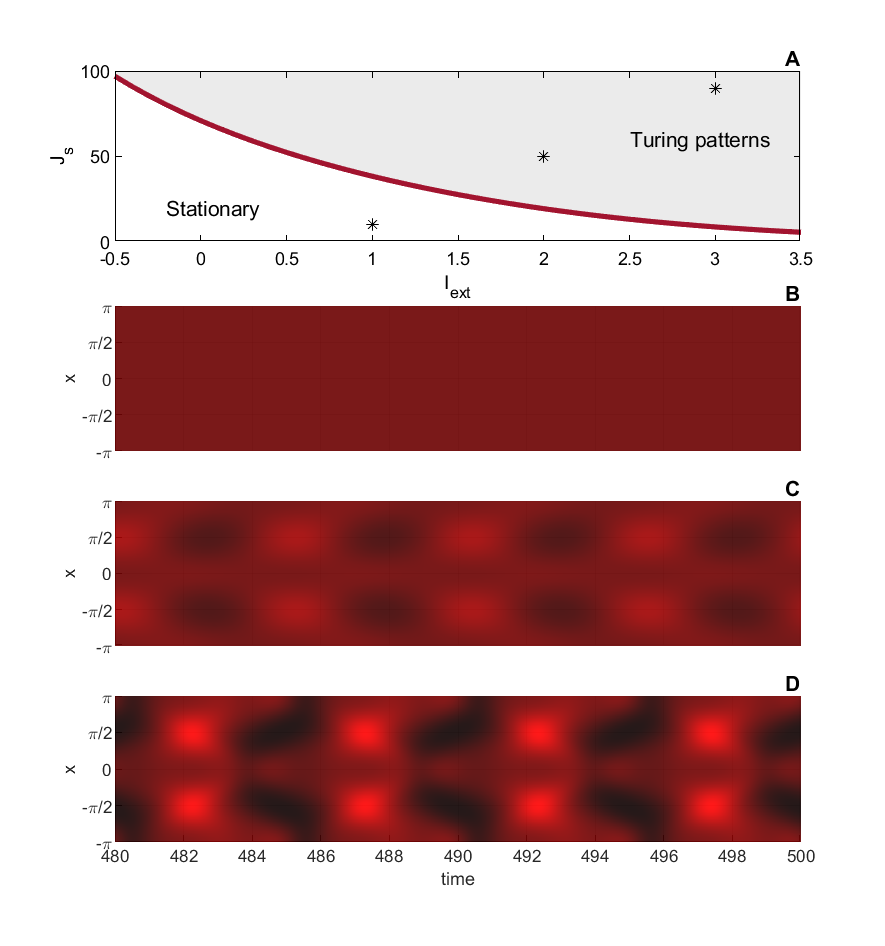}
   \caption{Pattern formation on the ring network. In the panel {\bf A}, the bifurcation line that separates the asynchronous domain from a Turing pattern regime, given by (\ref{CE}), is shown. The panels {\bf B}, {\bf C}, {\bf D} present the corresponding activities given by  (\ref{FR}) for the three sets of parameters depicted in panel {\bf A}, namely $(I_{ext}, J_s)=(1,10)$, $(I_{ext}, J_s)=(2,50)$ and $(I_{ext}, J_s)=(3,90)$. In panel {\bf B}, the convergence toward the steady state is observed. In panel {\bf C},  the onset of Turing patterns can be spotted, while in panel {\bf D}, the patterns in the activity are clearly defined.  Parameters: $\tau =5$, $T=5$, $\alpha=1/2$.  }
   \label{Figurea}
      \end{center}
\end{figure}

\begin{figure}[t!]
\begin{center}
    \includegraphics[width=\textwidth]{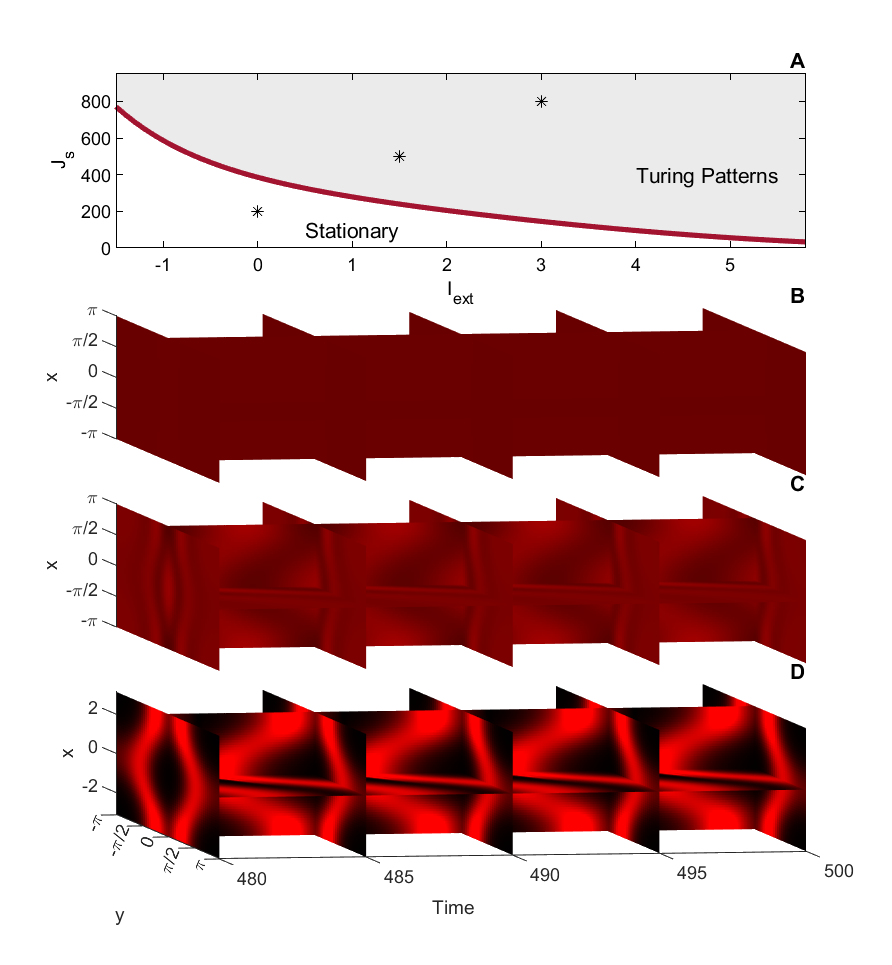}
       \caption{Pattern formation on the torus network. In the panel {\bf A}, the bifurcation line that separates the asynchronous state from a Turing pattern regime given by (\ref{CE}), is shown. The panels {\bf B}, {\bf C}, {\bf D} present the corresponding activities given by (\ref{FR}) for the three sets of parameters depicted in panel {\bf A}, namely $(I_{ext}, J_s)=(0,200)$, $(I_{ext}, J_s)=(1.5,500)$ and $(I_{ext}, J_s)=(3,800)$.  The convergence toward the steady state is illustrated in panel {\bf B}, while panels {\bf C} and {\bf D} display the onset of patterns of activity. Parameters: $T=5$, $\tau=5$, $\alpha=1/4.1$. See the supplementary material for different views of the figure. }
       \label{Figure6}
      \end{center}
\end{figure}

The last equation determines the stability of the asynchronous state. The time independent solution will be stable if all the eigenvalues $\lambda$ have negative real parts. We get in this way the so called the characteristic equation, formally written as 
\bd
\mathcal{C} (\lambda,k) = 0,
\ed
with
\bda
\mathcal{C} (\lambda,k) &=&      1-\hat J (k) \kappa_{\lambda}  \int_{0}^{+\infty}\frac{\partial S}{\partial h}(h_{\infty},r)q_{\infty}(r)\dd r -  \int_{0}^{+\infty}S_{\infty}(r) e^{- \int_0^{r}  S(h_{\infty},s)  +\lambda  \ddd s } \dd r \\
&&-\hat J (k) \kappa_{\lambda}      \int_{0}^{+\infty}S(h_{\infty},r) \int_0^{r} \frac{\partial S}{\partial h} (h_{\infty},a)q_{\infty}(a) e^{- \int_a^{r}  S(h_{\infty},s)  +\lambda  \ddd s } \dd a \dd r.
\eda
To characterize the crossing from stable solutions to oscillations, one needs then to find the roots of the characteristic equation, which in our case are  associated to the Fourier mode $k$ of $J$.  In all the simulations of the paper, the bifurcation line given by $\mathcal{C} (\lambda,k)=0$ was numerically solved for the maximum value of $\hat J(k)$, which is, as explained below, a function only of the distance $\sqrt{k_1^1+k_2^2}$.

In Fig.\ref{Figurea}, the representation of the line separating the stationary and Turing patterns regimes is depicted, along with the simulations of the corresponding activities extracted from the simulations of the mean field equation of three sets of parameters from both areas. The bifurcation line, with respect to parameters $I_{ext}$ and $J_s$, is represented in panel {\bf A}. In panels {\bf B}, {\bf C} and {\bf D}, respectively, the corresponding activity to the three sets of parameters shown in panel {\bf A} is plotted. Panel {\bf B} shows the convergence toward steady state, while in panel {\bf C}, one can already observe emergence of patterns. The patterns become more obvious when advancing in the Turing pattern area, as can be observed in panel {\bf D}.\\

In Fig. \ref{Figure6}, the same representation is made for the case $n=2$. Panel {\bf A} presents the line separating the stationary and Turing patterns regimes. In panel {\bf B} the stationary activity extracted from the mean field equation is shown, and the activity extracted from the same equation shows the onset of patterns in panels {\bf C} and {\bf D}.

One can notice, once again, the striking similarity between the representations of the neural activity given by the simulations of the full network, Fig. \ref{Figure1} and Fig. \ref{Figure7}, and the corresponding mean field representations - Fig. \ref{Figurea} and \ref{Figure6}, respectively,

\subsection{The Fourier Coefficient}
\begin{figure}[t!]
\begin{center}
    \includegraphics[width=\textwidth]{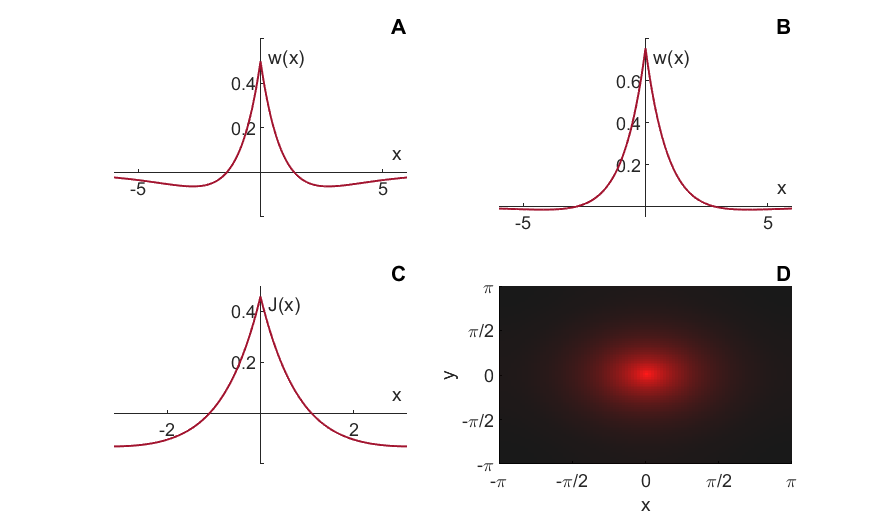}
       \caption{Kernel functions for the ring, respectively torus networks. Panel {\bf A} shows the graphical representation of the function $w(x)$ (with $\al=\frac{1}{2}$) used in the construction of the kernel $J$  in the case $n=1$. The corresponding kernel given by (\ref{Jkernel}) is represented in panel {\bf C}. The function  $w(x)$ (with $\al=\frac{1}{4.1}$) used for building the kernel in the case $n=2$ is depicted in panel {\bf B}, while the corresponding function $J$ is presented in panel {\bf D}.}
       \label{Figure11}
      \end{center}
\end{figure}

We consider in here a function $J$ that illustrates the case of rotationally symmetric and translationally invariant connections. We assume periodic domain, and consider $J$ defined on the domain $[-\pi,\pi]^n$, $n=1,2$.
Following the idea from \cite{Ermentrout_2}, Chapter 12, such a function can be constructed as
$$\begin{array}{ccccc}
J & : &  [-\pi,\pi]^n & \to &  \mathbb{R} \\
 & & x  & \mapsto & J(x)\\
\end{array}$$
where we define

\be\label{Jkernel}
J(x)=J_s\sum_{l_1=-\infty}^{+\infty}...\sum_{l_n=-\infty}^{+\infty} w\left(\left(\sum_{k=1}^n(x_k+2\pi l_k)^2\right)^\frac{1}{2}\right),
\ee 
for $x=(x_1,..,x_n)$, and $w$ is a scalar function 
$$\begin{array}{ccccc}
w & : &  \mathbb{R}  & \to &  \mathbb{R} \\
 & & t  & \mapsto & w(t)\\
\end{array}$$
with the property
\bd
\int_{-\infty}^{+\infty} w(t)\dd t < \infty.
\ed

The kernel $J(x)$ can be assessed numerically by sequentially iterating through the indices $l_i$. Because the function $w$ diminishes quickly, only a few iterations are needed. To guarantee a precise estimation of the kernel $J$, a tolerance level of $10^{-6}$ can be established (which is the case in our scenario). If adding another iteration doesn't noticeably change the values of $J$ beyond the tolerance level, the iteration is stopped, indicating a satisfactory approximation.

In this case one can compute the Fourier coefficient as
\be\label{formula}
 \hat J(k) := \int_{[-\pi,\pi]^n} J(y)e^{-i k\cdot y}\dd y  ,
\ee
where $k=(k_1,k_2,..,k_n)$ and $"\cdot"$ denotes the dot product in $	\mathbb{R} ^n$. Defining:
\be\label{formula1}
 l= \left(\sum_{i=1}^n k_i^2 \right)^{\frac{1}{2}},
\ee
we get
\bd
\hat J(k)= J_s \int_{\mathbb{R}} w(t)e^{-i l t}\dd t, \mbox{ for } n=1,
\ed
and 
\bd 
\hat J(k)= \bar J_s \int_{0}^\infty w(t) t J_0(t l)\dd t, \mbox{ for } n=2,
\ed 
where $J_0$ is the Bessel function of first kind of order zero and $\bar J_s = 2\pi J_s$. The last formula has been obtained 
from (\ref{formula}) by passing to the polar coordinates both in $x$ - space and $k$ -space and using the definition of $J_0$ function, see  \cite{baddour} for additional details.

Throughout the numerical simulations presented in this paper, we have considered a Mexican-hat type function
\bd
w(t)=e^{- |t|}- \al e^{- \frac{|t|}{2}},
\ed
where $\al$ is a constant. 
\\

In Fig. \ref{Figure11}, the choices of functions $w$ (panels {\bf A} and {\bf B}, for $n=1$, respectively $n=2)$, as well as the corresponding kernels obtained by the use of formula (\ref{Jkernel}) (panels {\bf C} and {\bf D}, for $n=1$, respectively $n=2)$)  used in all the simulations presented in this paper, are shown.\\

\noindent {\bf  Supplementary material: } For a better view of the two dimensional plots, we provide the Matlab figures Figure 2, Figure 6 and Figure 9 as a supplementary material.

\bibliographystyle{siamplain}
\bibliography{Pattern_Turing}
\end{document}